\documentclass[twocolumn,aps,pre]{revtex4}
\usepackage{color}
\usepackage{amsmath}    
\usepackage{amssymb}
\usepackage{graphicx}   
\usepackage{subfigure}
\usepackage{longtable}
\usepackage[utf8]{inputenc} %Codificacion utf-8
\newcommand{\ang}[1]{\langle #1\rangle}

\begin{document}

\title{Stochasticity enhances the gaining of bet-hedging \\ strategies in
  contact-process-like dynamics}

\author{Jorge Hidalgo$^{1}$, Simone Pigolotti$^{2}$, and Miguel A. Mu\~noz$^{1}$} 
\email{mamunoz@onsager.ugr.es}
\affiliation{$^{1}$ Departamento de Electromagnetismo y F\'isica de la
 Materia, and Instituto Carlos I de F\'isica Te\'orica y
  Computacional, Universidad de Granada, 18071 Granada, Spain.\\ 
  $^{2}$ Departament de Fisica i Enginyeria Nuclear, Universitat Politecnica
  de Catalunya, Rambla Sant Nebridi 22, 08222 Terrassa, Barcelona,
  Spain.}

\date{\today}
\begin{abstract}
  In biology and ecology, individuals or communities of individuals
  living in unpredictable environments often alternate between
  different evolutionary strategies to spread and reduce risks. Such
  behavior is commonly referred to as ``bet-hedging.'' Long-term
  survival probabilities and population sizes can be much enhanced by
  exploiting such hybrid strategies.  Here, we study the simplest
  possible birth-death stochastic model in which individuals can
  choose among a poor but safe strategy, a better but risky
  alternative, or a combination of both. We show analytically and
  computationally that the benefits derived from bet-hedging
  strategies are much stronger for higher environmental variabilities
  (large external noise) and/or for small spatial dimensions (large
  intrinsic noise). These circumstances are typically encountered by
  living systems, thus providing us with a possible justification for
  the ubiquitousness of bet-hedging in nature.
\end{abstract}


\maketitle

\section{Introduction}
In natural environments, individuals have to choose among a variety
of evolutionary strategies, characterized by different
payoffs and risks, which, in their turn, may change with
time. Particularly relevant is the case in which a choice is to be
made between a relatively safe strategy, with a low but stable payoff,
and a potentially very productive, but risky, variable
strategy. Micro-organisms able to metabolize two resources
\cite{Monod,Kuipers2008,Kuipers2011,Kuipers2014}, one of them
consistently available at a fixed though low level, and the second,
more abundant on average but fluctuating in time, constitute an
example of this.  In the absence of specific knowledge about
environmental conditions, individuals need to make a blind decision about
whether to specialize in consuming only one resource or, instead,
develop a hybrid ``bet-hedging'' strategy by alternating both options.
Similar forms of bet-hedging can also emerge in cases where limited
information from sensory systems is available
\cite{Seger,leibler}. Finally, bet-hedging can be exploited at a
community level --rather than on an individual basis-- by developing, for
example, phenotypic variability among individuals in a
population\cite{leibler,Kuipers2014}.

The concept of bet-hedging was first formalized in the context of
information theory \cite{kelly} and portfolio management
\cite{Fernholz}. Later, it was conjectured that living organisms may
employ bet-hedging strategies to decrease their risk in unpredictable
environments
\cite{Seger,Hastings,comins-hamilton-may,hamilton-may,jansen99}. This
idea has been empirically confirmed in bacterial and viral communities
\cite{stumpf-virus,wolf1,wolf2,Kuipers2008,Kuipers2011,Kuipers2014,sneppen-virus},
in insects \cite{hopper}, in seed-dispersal strategies developed by
plants \cite{childs,venable, levin}, and in a wealth of other examples
in population ecology, microbiology, and evolutionary biology
\cite{Hastings}.

Given their ubiquity, bet-hedging strategies have attracted a lot of
interest from the perspective of evolutionary game theory
\cite{smith,nowak}. An interesting and nontrivial result in this
context is the so-called Parrondo's paradox \cite{parrondo1999,
  parrondo2000}, in which an alternation of two ``losing''
strategies can lead to a ``winning'' one. However, most of the existing
theoretical studies of this effect, including applications in
population dynamics \cite{Hastings}, rely on mean-field analyses
describing well-mixed communities.

In this paper, we aim to extend previous approaches and to
  show that the competitive advantage of bet-hedging hybrid
  strategies is much stronger in the presence of highly variable
  environments and/or in low-dimensional systems, where the effect of
  fluctuations, i.e., demographic noise, is maximal and mean-field
  predictions do not hold.

For this, we study a minimal mathematical model of bet-hedging. It is
based on the physics of the contact process (CP)
\cite{marro-dickman,Munoz-Gseminar,hinrichsen-book}, but with the twist that
individuals can randomly choose between two strategies: one
characterized by a fixed reproduction rate and the other by fluctuating
environmental-dependent rates.  By combining analytical and
computational results, we conclude that bet-hedging benefits are
greatly enhanced in noisy low-dimensional environments such as the ones
that living systems typically inhabit and end by discussing the
relevance of our results for more realistic models of biological
populations.

\section{The model: Contact Process with hybrid dynamics}
\label{model}
\begin{figure}
\centering \includegraphics[width=\columnwidth]{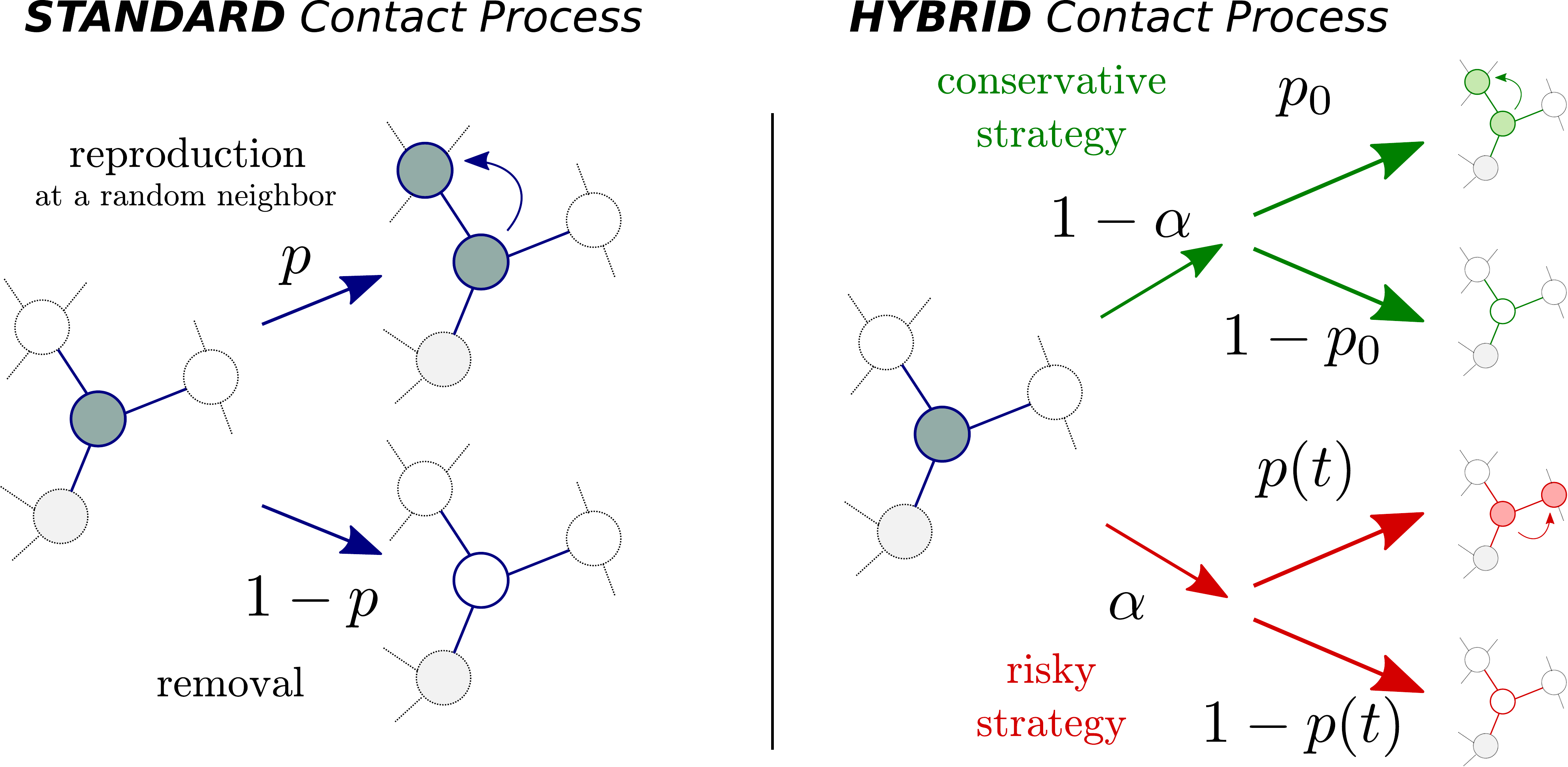}
\caption{Left: Sketch of the standard contact process dynamics: each
  occupied node (individual) in the lattice either (i) produces an
  offspring with probability $p$ at a random neighboring site
  (provided it is empty) or (ii) is removed from the lattice with
  complementary probability $1-p$. Right: Contact process with hybrid
  dynamics: at each time step, every individual chooses between the
  conservative (with probability $1-\alpha$) and the risky (with
  probability $\alpha$) spreading strategy. The conservative dynamics is
  characterized by a constant, relatively low spreading probability
  $p_0$, while the risky one depends on a stochastic variable $p(t)$,
  common to all particles in the system. }
\label{fig:cp_graphical}
\end{figure}

Our starting point is the simplest possible birth-death stochastic
model on a lattice, i.e., the CP
\cite{marro-dickman,Munoz-Gseminar,hinrichsen-book} (see Fig.~\ref{fig:cp_graphical},
left).  Individuals occupy a (square) lattice or network, with at most
one individual per site. At every discrete time step, each individual
can either produce (with probability $p$) an offspring at a randomly chosen
neighboring site --provided it is empty-- or die and be removed from
the system (with probability $1-p$). This simple dynamics can
--depending on the value of $p$-- either generate an active phase
characterized by a nonvanishing density of individuals or,
alternatively, lead ineluctably to the absorbing state in which the
population becomes extinct. A critical point, $p_c$, separates these
two distinct phases \cite{marro-dickman,Munoz-Gseminar,hinrichsen-book}.

We consider a variant of the CP dynamics in which individuals can
choose between two strategies (see Fig. \ref{fig:cp_graphical},
right): a ``conservative'' one, corresponding to exploitation of a
constantly available resource, and a ``risky'' one, exploiting a
variable/unpredictable resource. The conservative strategy corresponds
to a CP in which $p$ is kept constant at a relatively low value,
$p_0$. On the other hand, in the risky strategy, demographic
probabilities depend upon variable external conditions, i.e.,  $p =
p(t)$, where $p(t)$ is a random noise, common to all individuals in the
community. We focus on the simple case in which $p(t)$ is freshly drawn
at every (Monte Carlo) time step, and discuss later the case in which
the environment is temporally correlated.

Individuals can hedge their bets by randomly picking either of the two
competing strategies at each time step. This choice is controlled by the
``risk parameter'' $\alpha$: at each time step, each individual
independently adopts the risky strategy with probability $\alpha$ or the conservative one
with probability $1-\alpha$ . In the language of
game theory, $\alpha=0$ and $\alpha=1$ correspond to ``pure
strategies'' and the range $0<\alpha<1$ corresponds to a set of hybrid
strategies. In what follows, we assume that all individuals in
  the community are characterized by the same value of the risk
  parameter $\alpha$; variations in which $\alpha$ is
  individual-dependent are left for a future work. Key observables are
  the stationary density of individuals, $\rho$, the averaged
  (exponential) growth rate, $G$, and the mean extinction time of
  small populations, $\tau$ (see below).

\section{Theoretical insights}
\label{sec:maths}

In game theory, it is known that a hybrid plan consisting in the
alternation of two distinct pure strategies can outperform both of
them (see, e.g., \cite{kelly,parrondo1999,jansen99,Hastings}),
constituting an example of Parrondo's paradox. This effect plays
an important role for our aims in what follows.  In this section, we
discuss a simplified one-variable equation aimed at capturing the gist
of our model.

In particular, the simplifying assumptions we make here are as follows: (i) We
consider a mean-field limit in which spatial fluctuations are
neglected. (ii) We neglect nonlinear saturation terms; this is a valid
approximation only for low densities. (iii) We consider a continuous
time limit, as usually done to analyze the physics of the CP
and other particle systems
\cite{gardiner,marro-dickman,hinrichsen-book} (a discrete-time
  calculation is presented in Appendix A to prove the robustness of
  our conclusions against this assumption). In the continuous-time
limit, we consider the growth rate $p(t)=\bar p+\sigma\xi(t)$,
where $\bar p$ and $\sigma$ are constants --the mean and
  amplitude of the stochastic risky strategy $p(t)$, respectively--,
  and where $\xi(t)$ is a Gaussian noise with $\langle\xi(t)\rangle=0$
  and $\langle\xi(t)\xi(t')\rangle=\delta(t-t')$ [observe that even if
  we maintain the same notation as above, $p(t)$ and $p_0$ need to be
  interpreted as transition rates in the continuous-time
  approach]. The choice of a Gaussian probability distribution
  function for $p(t)$ enables us to obtain analytical calculations,
  but it has some ``technical'' drawbacks. In particular, being an
  unbounded distribution, $p(t)$ can take negative values; thus, in
  order to avoid interpretation problems, we need to restrict
  ourselves to the case $ 0 \ll \bar p -\sigma$ and $\bar p + \sigma \ll1$,
  where these effects should be negligible. In any case, even if
  specific details may depend on this choice, the general results and
  conclusions presented in what follows are robust against changes in
  this probability. This is explicitly illustrated in Appendix A for
  the case of uniform bounded distributions.

  Under these assumptions, the density of individuals $\rho(t)$ obeys
  the following rate equation:
  \begin{eqnarray}
  \dot \rho(t) & =
  \alpha \left[ p(t) \rho \left(1-\rho\right) - \left(1- p(t)\right) \rho \right] \nonumber\\
  & + (1-\alpha) \left[ p_0 \rho \left(1-\rho\right) -
    \left(1-p_0\right) \rho \right].
\label{eqm}
\end{eqnarray}
Defining the average spreading rate,
\begin{equation}
p_\mathrm{av}(\alpha) = \alpha \bar p + (1-\alpha)p_0,
\label{eq:pav}
\end{equation}
Eq. (\ref{eqm})  becomes
\begin{equation}
  \dot \rho(t) = \left(2 p_\mathrm{av}(\alpha)-1\right)\rho - 
p_\mathrm{av}(\alpha) \rho^2 + 2 \alpha \sigma \rho \left(1-\frac{\rho}{2}\right)\xi(t),
\label{eq:basic1}
  \end{equation}
  which, owing to the stochastic nature of $p(t)$, is a Langevin
  equation, to be interpreted in the It\^o sense.  Up to leading
  linear order, we have the approximation
\begin{equation}
\dot \rho \approx  \left(2 p_\mathrm{av}(\alpha)-1\right)\rho  + 2\sigma \alpha \rho  \xi(t),
\label{basic2}
\end{equation}
valid for $\rho\ll1$. Now changing variables (using It\^o calculus) to
$y=\log(\rho)$ and averaging over realizations $\langle\cdot\rangle$,
eq. (\ref{basic2}) becomes
\begin{equation}
 \frac{d}{dt}
\langle \log \rho \rangle =G(\alpha),
\end{equation}
where the sign of the {\it exponential growth rate},
\begin{equation}
  G(\alpha) = -2\sigma^2\alpha^2 + 2  p_\mathrm{av}(\alpha) - 1,
\label{parabolic}
\end{equation}
determines whether the population tends to shrink or [owing to the
absence of the nonlinear saturation terms in this approximation,
Eq. (\ref{basic2})] to grow unboundedly.  These two regimes are
separated by a critical point at which $G(\alpha)=0$. 

\begin{figure}
 \centering \includegraphics[width=0.9\columnwidth]{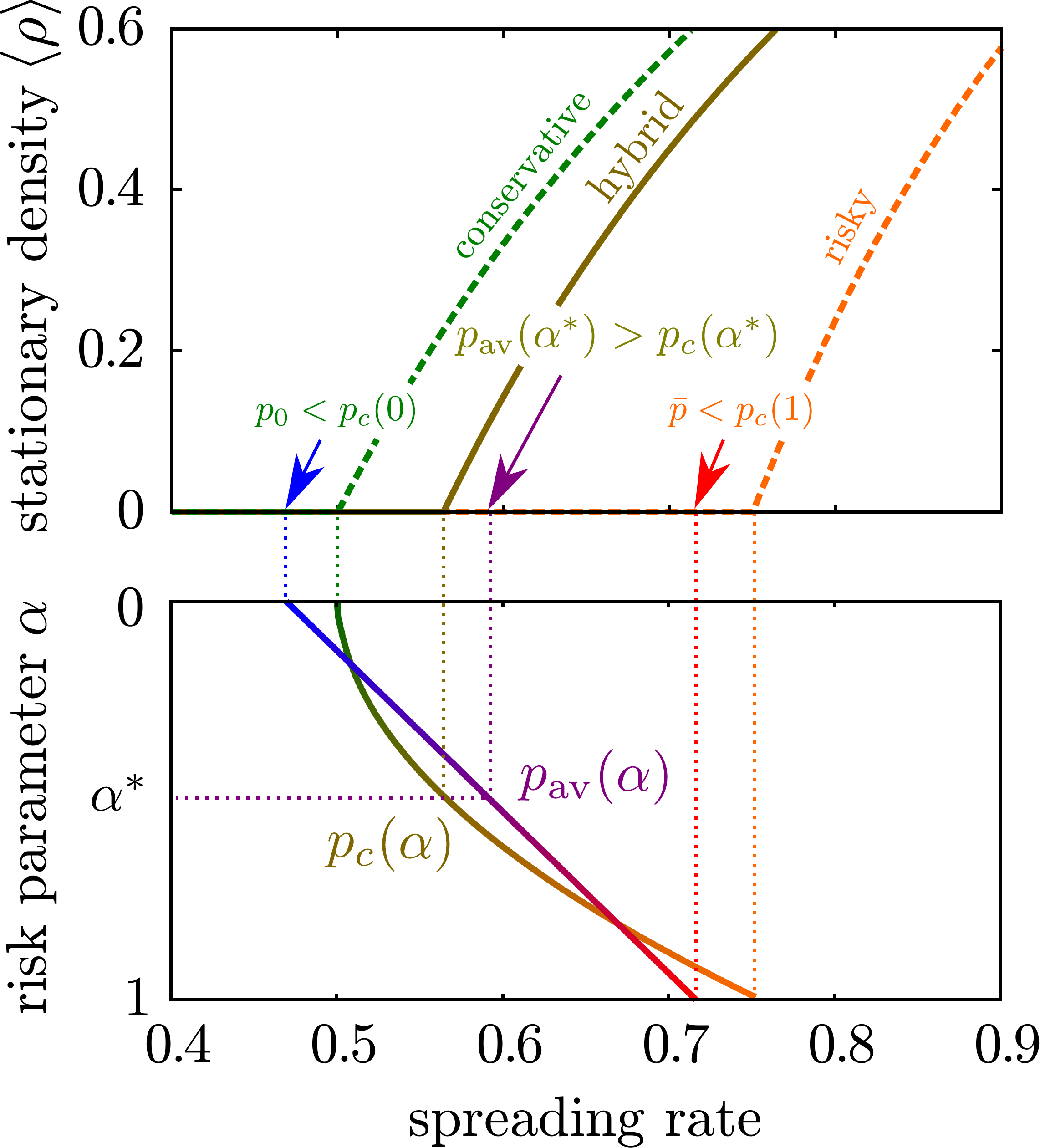}
 \caption{Top: Stationary density, computed via
     numerical integration of Eq. (\ref{eq:basic1}), for three types of
     strategies --conservative ($\alpha=0$), risky ($\alpha=1$), and a
     particular hybrid strategy ($0 < \alpha^* < 1$)-- as a function
     of the spreading probability, $p_0$ in the conservative case and
     $\bar{p}$ in the risky and hybrid cases. Critical points for
     the pure strategies can be obtained analytically:
     $p_c(\alpha=0)=1/2$ and $p_c(\alpha=1)=1/2 + \sigma^2$. Critical
     points for hybrid strategies lie between these two values. The
     lestmost (blue) and rightmost (red) arrows indicate a specific choice made for the two
     pure strategies, $\alpha=0$ and $\alpha=1$, respectively
     ($p_0=0.47$ and $\bar p=0.72$, with $\sigma^2=0.25$). Note
     that both of them are equally subcritical. The middle (purple) arrow marks
     the average spreading rate for a specific hybrid strategy (with
     $\alpha=\alpha^*$).  Bottom: Two lines are
     plotted as a function of $\alpha$: the effective value of the
     spreading rate $p_\mathrm{av}(\alpha)$ and the location of the
     critical point, $p_c(\alpha)$.  The first one interpolates
     linearly between our two pure-strategy choices (arrows) for $p_0$
     and $\bar{p}$ in the pure cases (as indicated by the blue-red
     color gradient). On the other hand, the second line is a
     quadratic interpolation between the two pure-strategy critical
     points (as indicated by the green-orange gradient).  As the two
     lines intersect each other, there exists a range of values of $\alpha$ for
     which $p_\mathrm{av}(\alpha) > p_c(\alpha)$ (supercritical) and others
     for which $p_\mathrm{av}(\alpha) < p_c(\alpha)$ (subcritical).  In
     particular, for intermediate values of $\alpha$, such as the
     marked $\alpha^*=0.5$, the stochastic alternation of two subcritical
     strategies results in a supercritical one. }
\label{fig:diagram}
\end{figure}

Keeping fixed all parameters but $\alpha$, we define the optimal strategy
$\alpha^* \in [0,1]$ as the one maximizing $G(\alpha)$.  This can
be either a pure or a hybrid strategy, depending on parameter
values. In particular, $\alpha^* =0$ for $\bar p < p_0$, $\alpha^* =1$
for $\bar p > p_0+2\sigma^2$, and $\alpha^*= {(\bar p -
  p_0)}/{2\sigma^2}$ for intermediate values of $\bar p$.  Observe
that the critical point is obtained for $ p_c(\alpha) = \frac{1}{2} +
\sigma^2\alpha^2$, i.e., the $\alpha$-dependent critical point
interpolates quadratically between the critical points of the pure
strategies, $p_c(0)=\frac{1}{2}$ and $p_c(1)=\frac{1}{2}+ \sigma^2$.
On the other hand, the average spreading rate $p_\mathrm{av}(\alpha)$
[Eq. (\ref{eq:pav})] is a linear-in-$\alpha$ interpolation between the
two limiting pure values.  

Figure \ref{fig:diagram} (top) shows the stationary density
  [obtained via numerical integration of Eq. (\ref{eq:basic1})] for
  the conservative, the risky, and an intermediate hybrid strategy.  The
  critical points at which the nontrivial steady states emerge
  coincide with the analytical predictions we have just made.  As
  explained in the caption to Fig. \ref{fig:diagram}, the different functional dependences for
  $p_\mathrm{av}(\alpha)$ and $p_c(\alpha)$ --linear and quadratic in
  $\alpha$, respectively-- enable the two curves to intersect each
  other, and thus, for intermediate values of $\alpha$ it is possible
  to have $p_\mathrm{av}(\alpha) > p_c(\alpha)$, i.e., a supercritical
  dynamics, even in the case (illustrated in Fig. \ref{fig:diagram})
  in which both pure strategies, $p_0$ and $\bar{p}$, are subcritical.
  Similarly, when the two pure strategies are active/supercritical,
  the same argument shows that a much higher stationary density can be
  achieved by hybrid strategies.

 This graphical representation --which we believe is new in the
    literature-- allows us to understand in a relatively simple and
    compact way the essence of Parrondo's paradox.  In what
    follows, we consider different types of pure strategies, either
    absorbing/subcritical or active/supercritical, and quantify the
    gain induced by bet-hedging in different settings, including fully
    connected (FC) networks (where the above mean-field approach should
    hold) and spatially explicit low-dimensional systems (where
    mean-field conclusions might break down).

\section{Computational results}

The calculation in the previous section provides valuable insight into
why hybrid strategies can be important, but it has some important
limitations. It is a mean-field calculation, thus neglecting spatial
structure. Moreover, Eq. (\ref{basic2}) includes only linear terms,
and thus it can only describe exponential growth starting from low
density rather than the steady-state behavior of the nonlinear
dynamics. To go beyond these limitations, here we perform direct Monte
Carlo simulations of the discrete model defined in Sec. \ref{model}
in large FC networks, and later we compare them with
similar simulations in one-dimensional (1D) two-dimensional (2D) and three-dimensional (3D) lattices. 

We implemented the CP dynamics using either synchronous/parallel or
asynchronous/sequential types of updatings. Here, we mostly focus on
the synchronous case.  In Appendix B, we show that asynchronous
updating leads to qualitatively similar results, even if quantitative
differences emerge. 
 
Simulations are initialized with a fully occupied configuration; then
the dynamics proceeds as follows: (i) At every step, a new value
of $p(t)$ is drawn from some probability distribution in $[0,1]$; in
most of this paper we use a truncated Gaussian $N(\bar p, \sigma^2)$
($\bar p$ and $\sigma^2$ are the mean and variance, respectively,
  of the nontruncated distribution) in which we fix possible values
$ p(t)<0$ to $p(t)=0$ and values $p(t)>1$ to $1$. This particular
choice may seem arbitrary, but we have verified that all the
forthcoming conclusions are robust and remain valid for, e.g., uniform
distributions. (ii) The network/lattice is updated
synchronously; with probabilities $\alpha$ and $1-\alpha$, each
individual selects the risky or the conservative strategy respectively.
(iii) Each individual either dies or reproduces with the
corresponding probabilities; all dying individuals are removed from
the system and afterward offspring are placed at random neighbors of
their corresponding parents, keeping the constraint of a maximum
occupancy of one individual per site (i.e., offspring trying to occupy
an already full site are simply removed).  Finally, (iv) time is
incremented in one unit and the process is iterated until a
  stationary state has been reached and steady-state measurements (of,
  e.g., $\rho$) are performed.

\begin{figure}
  \centering \includegraphics[width=\columnwidth]{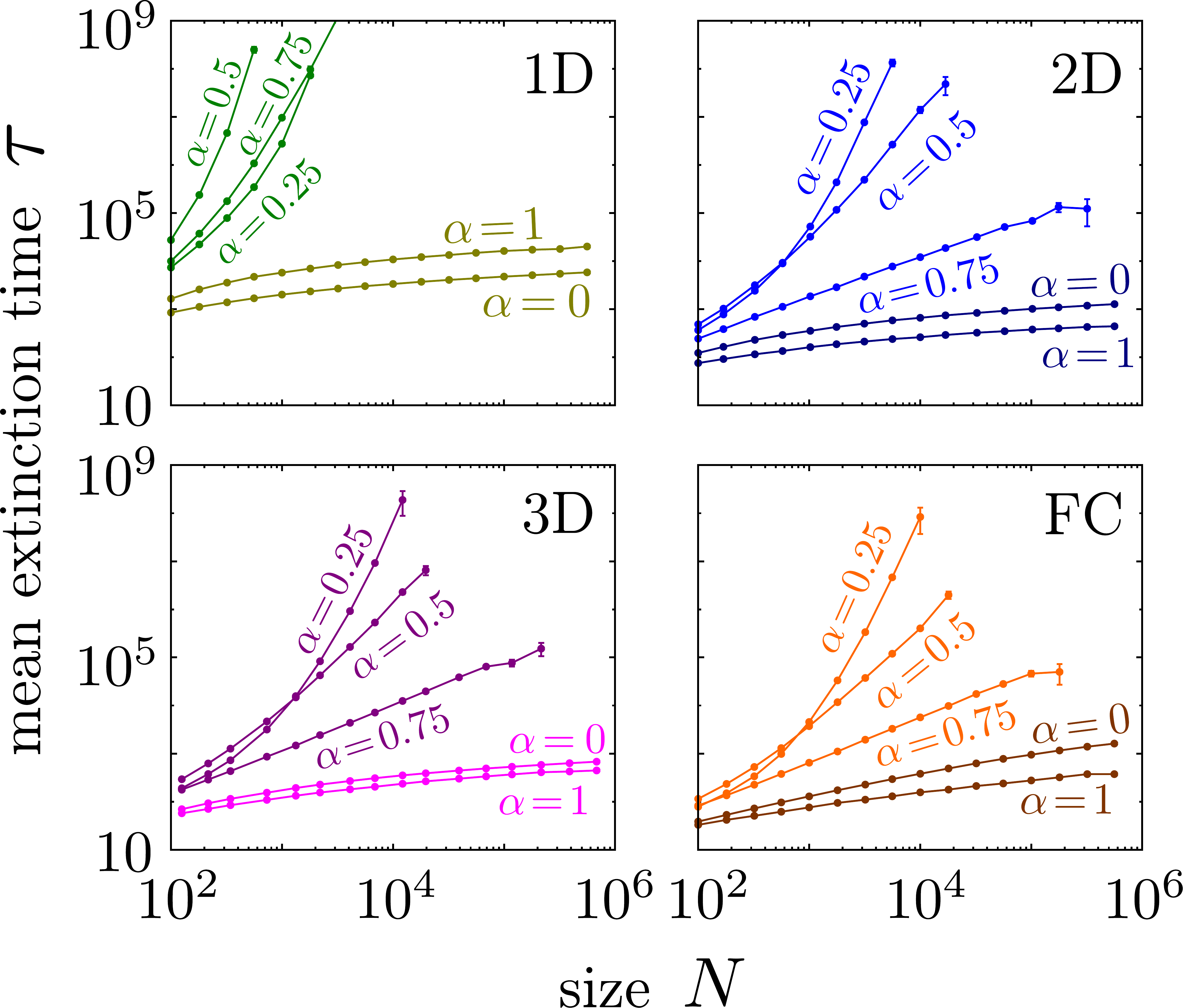}
 \caption{Mean extinction times as a function of system size $N$ for
   different strategies $\alpha$ and spatial dimensions. For our
   parameter choice, the two
   pure strategies, $\alpha=0,1$, have a logarithmic dependency
   (characteristic of subcritical behavior \cite{TGP}), while a
   range of hybrid strategies exhibits an
   exponential or power-law increase typical of active phases
   \cite{TGP}.  Parameter values $(p_0, \bar p, \sigma)$: 1D, $(0.71,
   0.80, 0.20)$; 2D, $(0.58, 0.71, 0.29)$; 3D, $(0.54,0.69,0.30)$, and
   fully connected (FC), $(0.499, 0.67, 0.33)$. Most error-bars are smaller than the symbol
   sizes.}
\label{fig:extinction-times}
\end{figure}

We begin by verifying the possibility of obtaining an active phase by
combining two strategies, each of them leading to an
absorbing/subcritical phase. To this aim, we fix the parameters $(p_0,
\bar p, \sigma)$ to poise the respective pure strategies ($\alpha=0$
or $\alpha= 1$) in the absorbing phase and study the behavior of
hybrid strategies at intermediate values of $\alpha$. To determine
whether or not a strategy leads to an active phase, we measure the
mean extinction time, $\tau$, as a function of the system size $N$.
Observe that, owing to fluctuations, any finite system is condemned to
end up in the absorbing state. However, its mean lifetime increases
exponentially with $N$, $\tau\sim\exp(N)$ in the active phase
\cite{gardiner}, making the system stable in the large-$N$
limit. Note that, in the presence of fluctuating parameters,
$\tau(N)$ can also scale as a power-law in the active phase
\cite{TGP}. On the other hand, a slow logarithmic increase,
$\tau\sim\log (N)$, is expected in the absorbing phase
\cite{gardiner,TGP}.  As shown in Fig. \ref{fig:extinction-times} for
different values of $\alpha$ and for different spatial dimensions,
$\tau$ grows logarithmically with $N$ for the two pure strategies
($\alpha=0,1$), as corresponds to both of them being absorbing, while
it increases exponentially for an intermediate range of hybrid
strategies, which depends upon the systems dimensionality. We
therefore conclude that in the CP the stochastic alternation of two
absorbing dynamics can lead to an active one, in agreement with
the linear-approximation above.

Some remarks are in order.  The advantageous consequences of
bet-hedging are not limited to the mean-field case, which can be
simply interpreted in terms of Eq. (\ref{eq:basic1}), but are
important also in low-dimensional systems where internal fluctuations
play a key role. Observe also that, as the phase boundaries depend on
dimensionality, different parameters are chosen for different panels in
Fig. \ref{fig:extinction-times}. We discuss later a way to
compare more clearly the strength of the effect as the system
dimensionality is changed. Finally, we have made no attempt here to
accurately determine the values of $\alpha$ delimiting the active
phase for each dimension, but have just confirmed the stabilizing effect of
hybrid strategies.

The goal now is to quantitatively analyze how the benefits of
bet-hedging depend on the level of stochasticity, both external
(environmental) and intrinsic (demographic).

\subsection{Environmental/external noise}
 
\begin{figure}
 \centering \includegraphics[width=0.9\columnwidth]{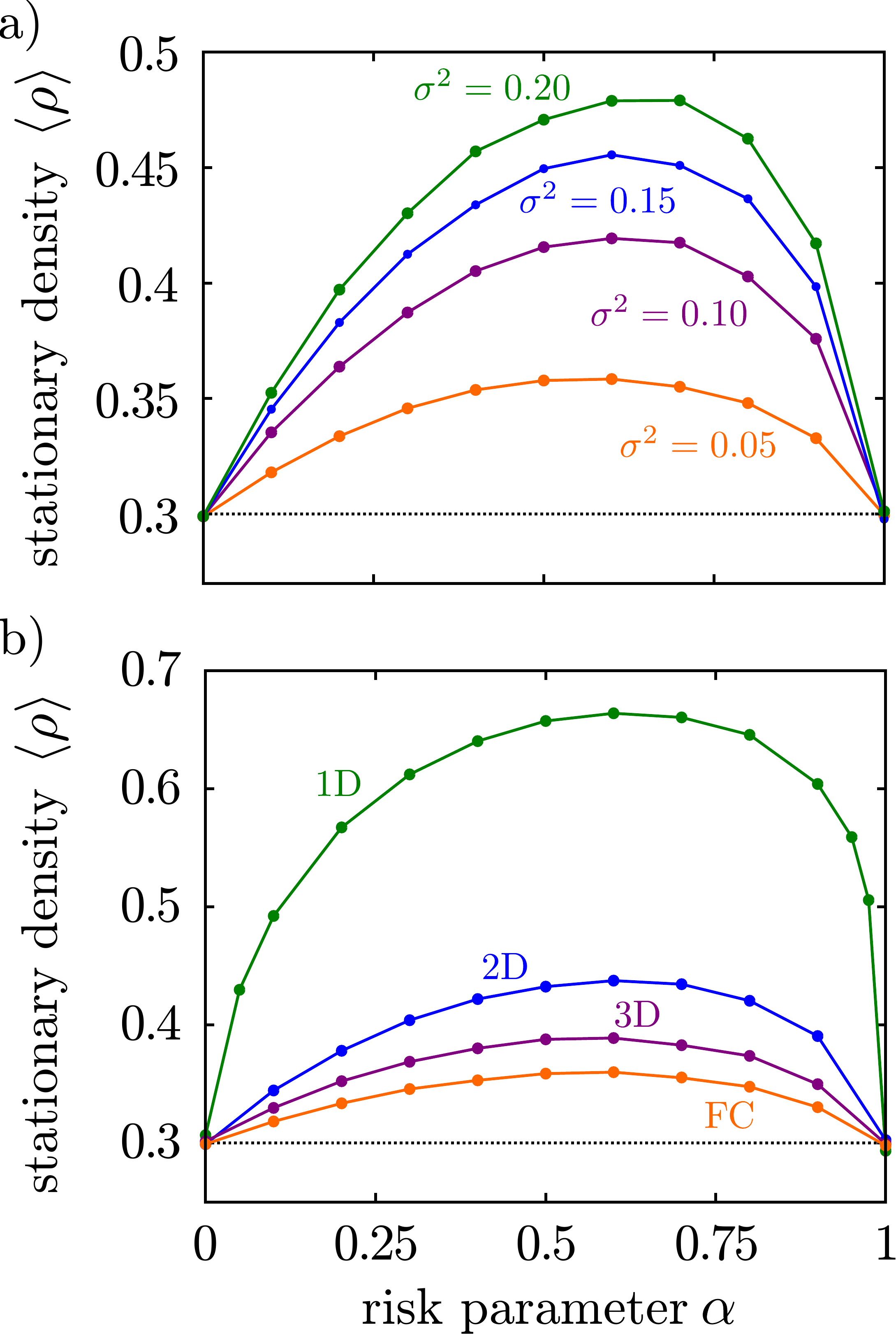} 
 \caption{(a) Effect of the external-noise variability ($\sigma^2$) on
   the stationary density for different bet-hedging strategies. Curves
   are results of Monte Carlo simulations of the fully-connected (FC) CP with
   bet-hedging. As $\sigma^2$ is increased, the optimal strategy
   induces higher stationary densities, even if the pure strategies
   $\alpha=0,1$ lead to the same density, $\langle
   \rho(\alpha=0,1) \rangle=0.3$. Parameter values are $N=10^4$,
   $p_0=0.567$. $(\bar p,\sigma^2)$ are $(0.628,0.05)$,
   $(0.699,0.10)$, $(0.765,0.15)$ and $(0.825,0.20)$ in the different curves.
   (b) Effect of dimensionality at fixed $\sigma^2$.
   The net benefit of bet-hedging is much enhanced at
   lower-dimensions. Parameters: $\langle \rho(\alpha=0,1)\rangle=0.3$,
   $\sigma^2=0.05$, $N=10^4$ for 1D, 2D and FC, $N=10648$ for 3D, and
   $(p_0,\bar p)$ are 1D:$(0.722,0.847)$, 2D:$(0.618, 0.704)$,
   3D:$(0.594,0.665)$ and FC:$(0.567,0.628)$. Error-bars are smaller
   than symbol sizes in all cases. }
\label{fig:2noises}
\end{figure}

First, we study the dependence on environmental variability
$\sigma^2$. To ease comparison, for each value of $\sigma^2$, we fix
the two pure strategies to have the same steady-state density,
$\langle\rho(\alpha=0)\rangle=\langle\rho(\alpha=1)\rangle = 0.3$, by
an appropriate choice of the only remaining free parameters, $p_0$ and
$\bar{p}$, respectively. Observe that, at variance with the
  previous section, here the pure strategies are taken to be
  ``equally'' active (same steady-state density), but we could have
  also taken them to be equally absorbing (same extinction
  time). The reason for this choice is that it allows for a much faster and easier
  computational implementation.  We then analyze how the steady-state
averaged density $\rho$ depends on $\alpha$ for different values of
$\sigma^2$.  Figure \ref{fig:2noises}(a) clearly illustrates that, in the
case of FC (mean-field) lattices, more variable
environments allow bet-hedging strategies to achieve much higher
stationary densities.  The same trend holds for low-dimensional
lattices (not shown): the larger the external noise, the more benefits
a community can derive from conveniently exploiting bet-hedging.

This observation is consistent with the linear analysis embodied in
Eq.(\ref{basic2}). Using the definition of $G(\alpha)$ and keeping the
environmental variance $\sigma^2$ as a control parameter, $p_0$ and
$\bar p$ can be fixed by imposing identical growth rates for the pure
strategies, $G(0)=G(1)\equiv G_{0,1}$. Under this constraint, the
maximum possible growing rate is
\begin{equation}
  \max (G) = G(\alpha=1/2) = G_{0,1} + \frac{\sigma^2}{2},
\label{max}
\end{equation}
predicting a linear increase in the optimal G with $\sigma^2$. 

As a final remark, observe that, although the two pure strategies
  have been set to be equivalent (in the sense that both lead to the
  same stationary density), the optimal strategy $\alpha^*$ in
  Fig. \ref{fig:2noises}(a) (maximizing $\langle \rho \rangle$) tends to
  be slightly larger than that provided by the approximate analytical
  prediction $\alpha^*=1/2$ (maximizing $G$).  We have checked that
  this ``favoring'' of the risky strategy strongly depends on the
  details of the implementation, as we have not observed it with
  asynchronous updating in the dynamics (see Appendix B).  So far, we
  have not attempted to determine the optimal strategy $\alpha^*$ for
  each case, but just to confirm the gain enhancement of bet-hedging
  in the presence of larger fluctuations.

\subsection{Dimensionality and demographic/intrinsic noise}

A main feature of low-dimensional models in statistical mechanics is
that intrinsic fluctuations (or {\em demographic} stochasticity, in
the language of population dynamics) play a more dramatic role than they
do in high dimensions \cite{Binney}, where they can be safely
neglected in mean-field-like approximations. We assume --and then
explicitly verify-- that smaller spatial dimensions effectively
correspond to larger levels of demographic noise.

We now explore the effect of dimensionality on bet-hedging; the
spatial dimension of the systems is varied while keeping a fixed
external noise variance $\sigma^2$. As above, to ease comparison, we
choose pure strategies for each dimension so that $\langle \rho(0)
\rangle= \langle \rho(1) \rangle =0.3$ (i.e., both pure strategies are
equally active) and measure computationally $\langle\rho\rangle$ as a
function of $\alpha$ for hybrid strategies in each dimension.

Figure \ref{fig:2noises}(b) clearly illustrates that the benefits of
bet-hedging are much enhanced as the system dimensionality is
decreased, allowing for much higher densities. In particular,
1D systems can accommodate twice as much density as
FC (infinite-dimensional) lattices.

A simple mathematical argument allows us to qualitatively --even if
not quantitatively-- understand this finding. Demographic noise is the
key ingredient, missing in the mean-field limit. Therefore, we
generalize Eq. (\ref{basic2}) to include a demographic-noise term of
tunable amplitude $\gamma$ \cite{gardiner,nature} as well as the
above-neglected leading nonlinear term
\begin{equation}
  \dot \rho(t) \approx
  \left(2p_\mathrm{av}(\alpha) -1 \right) \rho -
  p_\mathrm{av}(\alpha) \rho^2 +
  2\alpha\sigma\rho \xi(t) + \gamma \sqrt{\rho} \eta(t),
\label{basic-noise}
\end{equation}
where $\eta(t)$ is a Gaussian white noise. As usual, the square-root
factor multiplying the noise amplitude of demographic fluctuations is
a direct consequence of the central limit theorem, which, in particular,
imposes that fluctuations disappear in the absence of activity ($\rho=0$) \cite{gardiner}.

Equivalently to Eq.~(\ref{basic-noise}), we can write down the
  Fokker-Planck equation for the probability distribution $P(\rho,t)$
  \cite{gardiner}.  To work in the quasi-stationary approximation
  \cite{gardiner,Dickman,nature} (i.e., to avoid technical problems
  stemming from the existence of an absorbing state at $\rho=0$), we
  include a small and constant drift $\varepsilon$, which is a
  constant added on the right hand side of Eq. (\ref{basic-noise}), giving:
\begin{eqnarray}
\partial_t P(\rho,t) &=& -\frac{\partial}{\partial \rho}\left[\left(\varepsilon+\left(2p_\mathrm{av}(\alpha) -1 \right) \rho - p_\mathrm{av}(\alpha) \rho^2\right)P(\rho,t)
\right] \nonumber\\
&&+\frac{1}{2}\frac{\partial^2}{\partial \rho^2}\left[(\beta^2\rho+\gamma^2)\rho P(\rho,t)\right],
\end{eqnarray}
where, for convenience, we have introduced $\beta=2 \alpha \sigma$.
The associated stationary probability distribution function then reads:
\begin{equation}
 P_\mathrm{st}(\rho) = \left\{
 \begin{array}{l}
   C_1\, \rho^{\frac{2\varepsilon}{\gamma^2}-1} \exp{\left( \frac{2 (2 p_\mathrm{av}-1) \rho - p_\mathrm{av} \rho^2}{\gamma^2} \right)},  \quad \beta=0\\\\
   C_2\, \rho^{\frac{2\varepsilon}{\gamma^2}} \left(\gamma ^2+\beta ^2 \rho\right)^{\frac{2(2 p_\mathrm{av}-1)}{\beta^2}+\frac{2p_\mathrm{av}\gamma^2}{\beta^4} -\frac{2\varepsilon}{\gamma^2} -1} \times\\
  \qquad \exp{\left(-\frac{2 p_\mathrm{av} \rho}{\beta ^2}\right)} ,\quad \beta>0,
 \end{array}
 \right.
 \label{eq:fp-demographic-st}
 \end{equation}
 where $C_1$ and $C_2$ are normalization constants. From this equation
 we can compute the averaged quasi-stationary density 
\begin{equation}
  \langle \rho \rangle = \int_{0^+}^\infty d\rho \, \rho \,
  P_\mathrm{st}(\rho)
\label{density}
\end{equation}
as a function of parameter values.
 \begin{figure}
 \centering  \includegraphics[width=0.9\columnwidth]{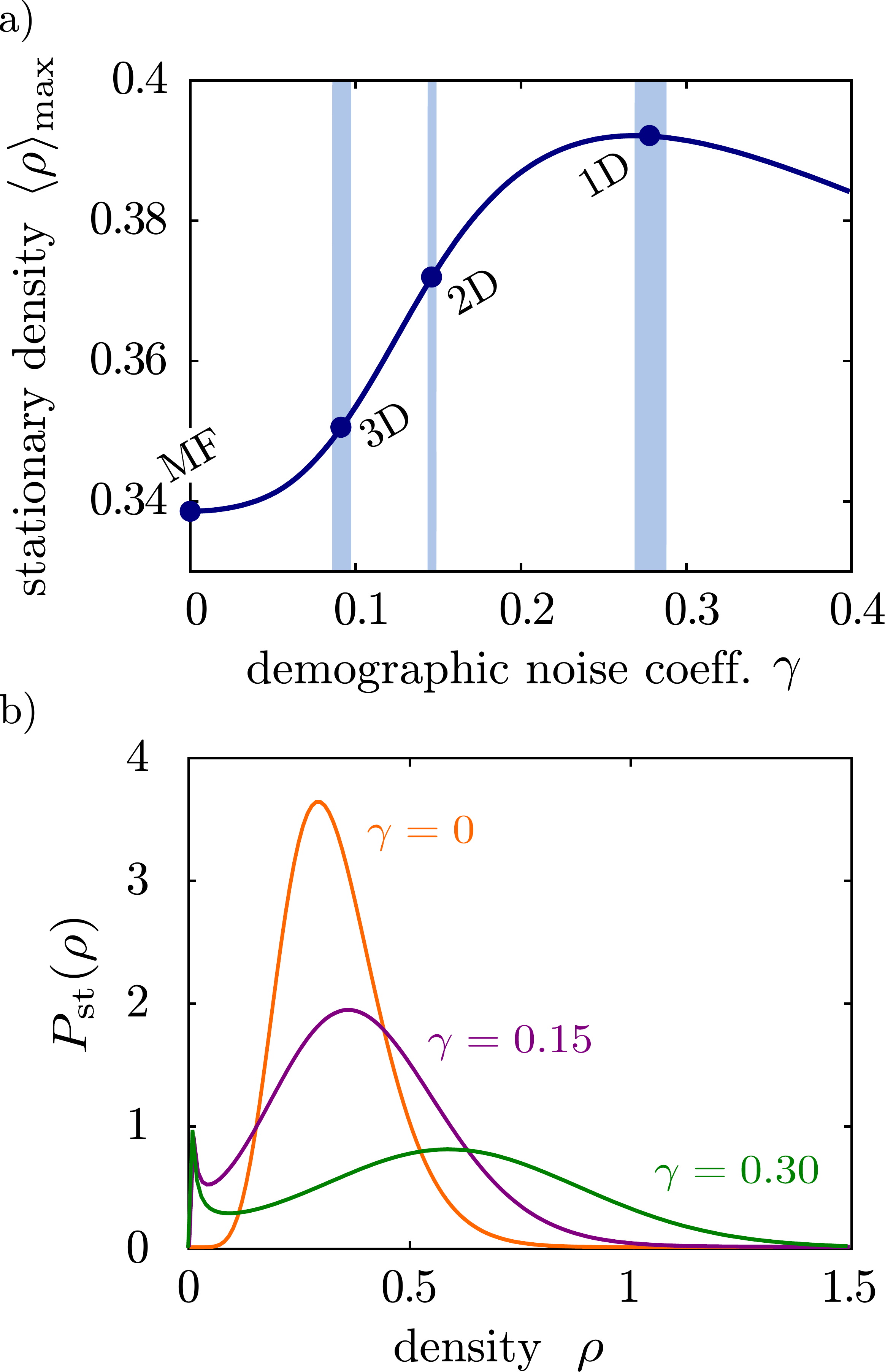}
 \caption{(a) Stationary density of the optimal hybrid strategy as a
   function of the demographic noise amplitude $\gamma$: each point
   was computed using the value of $\gamma$ inferred from
   Eq. (\ref{eq:fp-demographic-st}), for a different spatial dimension
   ($1$, $2$, and $3$), and $\gamma=0$ for the mean-field case.  Parameters $p_0$, $\bar p$, and $\sigma$ are the
   same as in Fig. \ref{fig:2noises}(b) and $\gamma$ is tuned to produce
   $\ang{\rho(\alpha=0)}=\ang{\rho(\alpha=1)}=0.3$. The inferred
   $\gamma$ changes slightly for $\alpha=0$ and $\alpha=1$ as
   reflected in the error bars (shaded region). These results confirm
   that the effective noise amplitude $\gamma$ increases as the system
   dimensionality decreases and that the benefits of bet-hedging are
   enhanced by demographic noise. However, the curve becomes nonmonotonous for larger values of $\gamma$. (b) Density distributions
   in the optimal strategy $\alpha^*$ for different demographic noise
   amplitudes $\gamma$: We represent the quasistationary solution
   of Eq. (\ref{eq:fp-demographic-st}) ($\varepsilon=10^{-3}$). The
   calculus fails for higher values of $\gamma$, as the probability of
   decay into the absorbing state (emerging peak at $\rho=0$) becomes
   non-negligible. The functional dependence of $\langle
     \rho\rangle_\mathrm{max}(\gamma)$ does not change qualitatively
     for other choices of this parameter, such as
     $\varepsilon=10^{-2}, 10^{-3}$ or $10^{-4}$.} 
 \label{fig:demographic}
 \end{figure}
The effective dimension-dependent value of $\gamma$ can now be
 inferred from the condition that --fixing the remaining parameters
 (i.e., $p_0$, $\bar p$, and $\sigma^2$) as in each of the spatially
 explicit simulations [see caption of Fig. \ref{fig:2noises}b]-- the
 quasi-stationary density in Eq. (\ref{density}) satisfies $\langle
 \rho(\alpha=0) \rangle=\langle \rho(\alpha=1) \rangle=0.3$.  The
 resulting values of $\gamma$ are $\gamma= 0.09, 0.15$, and $0.28$, for
 dimensions $3$, $2$, and $1$, respectively (each value is the average
 of two very close results obtained for the two pure strategies),
 confirming that, as expected, $\gamma$ --effectively representing the
 amplitude of demographic noise-- increases upon lowering the spatial
 dimension.

 Having determined, for each dimension, the level of demographic
 fluctuations $\gamma$, we now use Eqs. (\ref{eq:fp-demographic-st})
 and (\ref{density}) to compute the maximum density as a function of
 $\alpha$.  Results are shown in Fig. \ref{fig:demographic}, which
 reveals that the benefits of bet-hedging are enhanced for larger
 demographic noises and, thus, for lower spatial dimensions.

 We remark that this phenomenological single-variable theory only
 provides a qualitative explanation of the phenomenon and does not
 quantitatively reproduce the actual stationary densities in Fig.
 \ref{fig:2noises}(b).  Observe also that for very high noise
 amplitudes the curve in Fig. \ref{fig:demographic} veers down, while
 this effect is not seen when reducing the system dimensionality.  A
 more rigorous analytical approach to this problem --including the
 explicitly spatial dependence in Eq. (\ref{basic-noise})-- is a
 challenging task, beyond the scope of the present work.

\subsection{Time-correlated environments}

In the model we have discussed, the timescale of environmental changes is
the same as the generation times cale. However, real biological
populations have to cope with environmental conditions varying on
time scales possibly longer \cite{red-noise} than the individual generation
time. To address this important generalization, we checked how our
main results change upon varying the temporal correlation of the state
of the environment.  In particular, we simply described $p(t)$ as an
Ornstein-Uhlenbeck process (see, e.g., \cite{gardiner}) of average
$\bar{p}$ and variance $\sigma^2$ and study the effect of varying its
correlation time. 

Detailed results of this study are summarized in Appendix C. Our main
conclusion, i.e., that the benefits of bet-hedging strategies are
enhanced in lower-dimensional systems, remains unaltered. In addition,
considering an environment correlated over a few generations enhances
the advantage of bet-hedging in all dimensions, although this effect
is significantly stronger in low dimensions. Finally, the optimal
strategy becomes more conservative for environments characterized by a
very long correlation time. These results can be intuitively
understood by thinking that, if the environment is persistently
unfavorable for a long time, the extinction risk is very high, and
bet-hedging strategies become more crucial for survival. A much
  more detailed analytical characterization of bet-hedging dynamics
  under correlated environments is left for a future study.

\section{Conclusions} 
Summarizing, our main finding is that the relative benefit of
developing bet-hedging strategies is strongly enhanced in
highly fluctuating low-dimensional systems, where both internal and
external sources of variability greatly foster dynamical fluctuations,
leading to a strong departure from mean-field behavior.  Given that
these conditions are often met by biological populations --as for
instance, in bacterial colonies competing at the front of a range
expansion in noisy environments
\cite{Ben-Jacob,Review-Nelson,weber2014chemical}-- our results support
the importance of bet-hedging in nature.  This
being said, of course, more realistic models --including some
realistic ingredients such as, for example, the possibility of
``dormant'' states and not just birth and death processes-- would be
required to approach viral or bacterial communities and their
bet-hedging more closely.

The kind of trade-off considered in this paper, between a stable and a
fluctuating resource, is possibly the simplest example of bet-hedging,
both biologically relevant and natural to understand using tools of
nonequilibrium statistical physics.  However, we conjecture that the
increased strength of bet-hedging in low dimensionality is a general
phenomenon, present in other recently studied examples of trade-offs,
for example, between reproduction and motility
\cite{Frey2014,pigolotti2014selective} and in pairwise games
\cite{rulands2014specialization}.

Our preliminary results presented in Appendix C show that the
  effect described in this paper is still present in correlated
  environments.
  However, for very long correlation times, bet-hedging  strategies are 
  disfavored compared to short-correlated environments, but they always
  provide an advantage with respect to pure strategies. In view of
  these
 preliminary results, it will be of interest to investigate from a
  general perspective and in more depth the influence of
  environmental-noise temporal autocorrelations on bet-hedging, as
well as the difference between exploiting bet-hedging individually and exploiting it
at a community level. We believe that this work will provide a
physical framework to answer these and similar challenging questions
which might be of interest in biology and ecology.

\appendix

\section{Uniformly distributed environment}
In this Appendix we study the case in which the spreading
  probability for the risky strategy, $p(t)$, is bounded and uniformly
  distributed in the range $[\bar p-\delta, \bar p + \delta]$, where
  the parameter $\delta$ encapsulates the level of environmental
  variability. In particular, to avoid negative values, we take
  $\delta<\bar p$ and $\delta<1-\bar p$.  This type of distribution
  allows for analytical treatment using a discrete-time approach,
  which is common in the study of game theory \cite{kelly}. To
  proceed, we take the linearized rate equation derived in the 
  text, Eq. (\ref{basic2}), and write it in a discrete-time form
  (using one-unit time steps) and replace $p(t)$ with its explicit form for the
  uniform distribution,
\begin{equation}
\rho(t) = 2 p_\mathrm{av}(\alpha) \rho(t-1) + 2\alpha\delta \rho(t-1) u(t-1),
\end{equation}
where $u(t)$ is a uniformly distributed variable in the range
$[-1,1]$, for any integer $t \ge0$.

Assuming an initial density $\rho(0)$ and discretizing the range
of values of $u$, $u=(u_1,...,u_{U})$, the previous equation becomes
\begin{equation}
\rho(t) = \prod_{i=1}^{U} \left[ 2 p_\mathrm{av}(\alpha) + 2\alpha\delta u_i \right]^{n_i} \rho(0),
\end{equation}
where $n_i$ is the number of times that a value $u_i$ is obtained, and
therefore, $\sum_i n_i = t$.  The exponential growth rate is derived
from its discrete form, $G=\lim_{t\rightarrow\infty} \frac{1}{t}
\log\frac{\rho(t)}{\rho(0)}$ \cite{kelly},
\begin{equation}
G(\alpha) = \sum_{i=1}^U \frac{n_i}{t}\log \left[2 p_\mathrm{av}(\alpha) + 2\alpha\delta u_i\right], 
\end{equation}
which in the continuum limit becomes
\begin{eqnarray}
G(\alpha) &=& \int_{-1}^{1} du \frac{1}{2\delta} \log \left[2
  p_\mathrm{av}(\alpha) +  
2\alpha\delta u \right] = \log(2)-1 +\nonumber\\
&& \frac{1}{2\delta\alpha}\log{\frac{\left(p_\mathrm{av}(\alpha)+\alpha\delta\right)^{p_\mathrm{av}(\alpha)+\alpha\delta}}{\left(p_\mathrm{av}(\alpha)-\alpha\delta\right)^{p_\mathrm{av}(\alpha)-\alpha\delta}}},
\label{eq:Guniform}
\end{eqnarray}
for any $\alpha\in(0,1]$, and $G(0)=\log\left(2p_0\right)$ for $\alpha=0$.

\begin{figure}[t]
\centering  \includegraphics[width=\columnwidth]{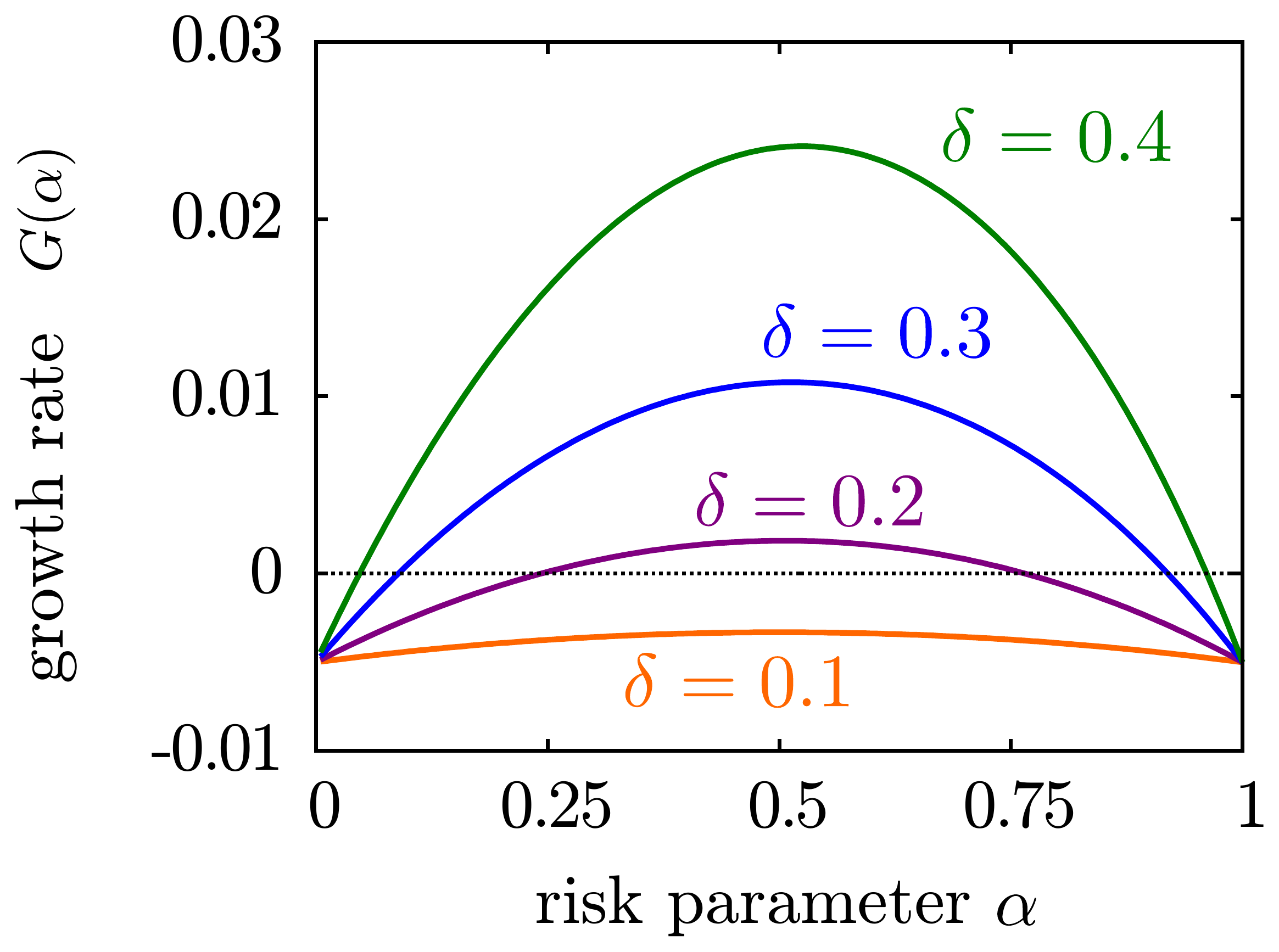}
\caption{Exponential growth rate $G$ as a function of the risk parameter
  $\alpha$ as given by Eq. (\ref{eq:Guniform}) for uniformly
  distributed environments: $p(t)=\bar p + \delta u(t)$, where $\bar
  p$ and $\delta$ are constants and $u(t)$ is a uniform noise
  distributed in $[-1,1]$. Different curves correspond to
  different values of the parameter $\delta$; in all cases the
  corresponding pure strategies have been tuned to give
  $G(\alpha=0)=G(\alpha=1)=-0.005$ in the absorbing/subcritical phase.
  The optimal strategy is always a hybrid one, very close to
  $\alpha^*=0.5$. In particular, for $\delta\gtrsim0.15$,
  $G(\alpha^*)>0$, allowing for active/supercritical dynamics.
  Furthermore, the maximum growth rate for this strategy increases
  with the amplitude of the environmental noise. Parameter values:
  $p_0=0.498$ fixed in all cases, and $(\bar p,
  \delta)=(0.501, 0.1), (0.511, 0.2), (0.528, 0.3)$ and $(0.553, 0.4)$. Note
  that $\delta<\bar p$ and $\delta < 1-\bar p$ in all cases.}
\label{fig:Guniform}
\end{figure}

Figure \ref{fig:Guniform} shows the solution $G$ as a function of
  $\alpha$ [Eq. (\ref{eq:Guniform})] for different choices of the
  environmental variability $\delta$. In this example, we have tuned the
  parameters $p_0$ and $\bar p$ to be equally subcritical, i.e.,
  $G(0)=G(1)<0$, but similar curves can be obtained if
  $G(0)=G(1)>0$. We see that (i) the optimal strategy, in the
  sense of having a maximal $G$, always lies at intermediate values
  of $\alpha$; (ii) the growth rate for such optimal strategy
  increases with the amplitude of fluctuations $\delta$; and (iii)
  for sufficiently large values of $\delta$, a combination of
  two subcritical strategies gives rise to a supercritical one, as
  $G(\alpha^*)>0$ for $\delta\gtrsim0.15$.  Moreover, we have tested
  these results in Monte Carlo simulations, as well as with different
  lattice dimensions, obtaining plots similar to
  Fig. \ref{fig:2noises} and
  Fig. \ref{fig:2noises-asynchronous}. Summarizing, in the case in
  which $p(t)$ is uniformly distributed, the same conclusion obtained
  for Gaussian distributions holds: the benefits of bet-hedging are
  stronger in the presence of both extrinsic and intrinsic
  fluctuations.

\section{Model with asynchronous updating}

In this Appendix, we verify the robustness of our results when an
asynchronous-updating version of the CP
\cite{marro-dickman} is implemented. At each time step, one of the
existing $N_\mathrm{act}(t)$ active particles is randomly selected;
with probability $\alpha$, the particle chooses the risky strategy or,
with complementary probability $1-\alpha$, the conservative one.  As
above, in the first case, it reproduces with probability $p(t)$ or
dies with probability $1-p(t)$, where $p(t)$ changes with the
environment, while for the conservative strategy it reproduces with
probability $p_0$ or dies with probability $1-p_0$. Time is
incremented in $1/N_\mathrm{act}(t)$. After all particles in the
network have been updated once on average (i.e. after time increases
in one unit) another value of $p(t)$ is drawn from a Gaussian
distribution $N(\bar p, \sigma^2)$.

\begin{figure}[tb]
 \centering \includegraphics[width=\columnwidth]{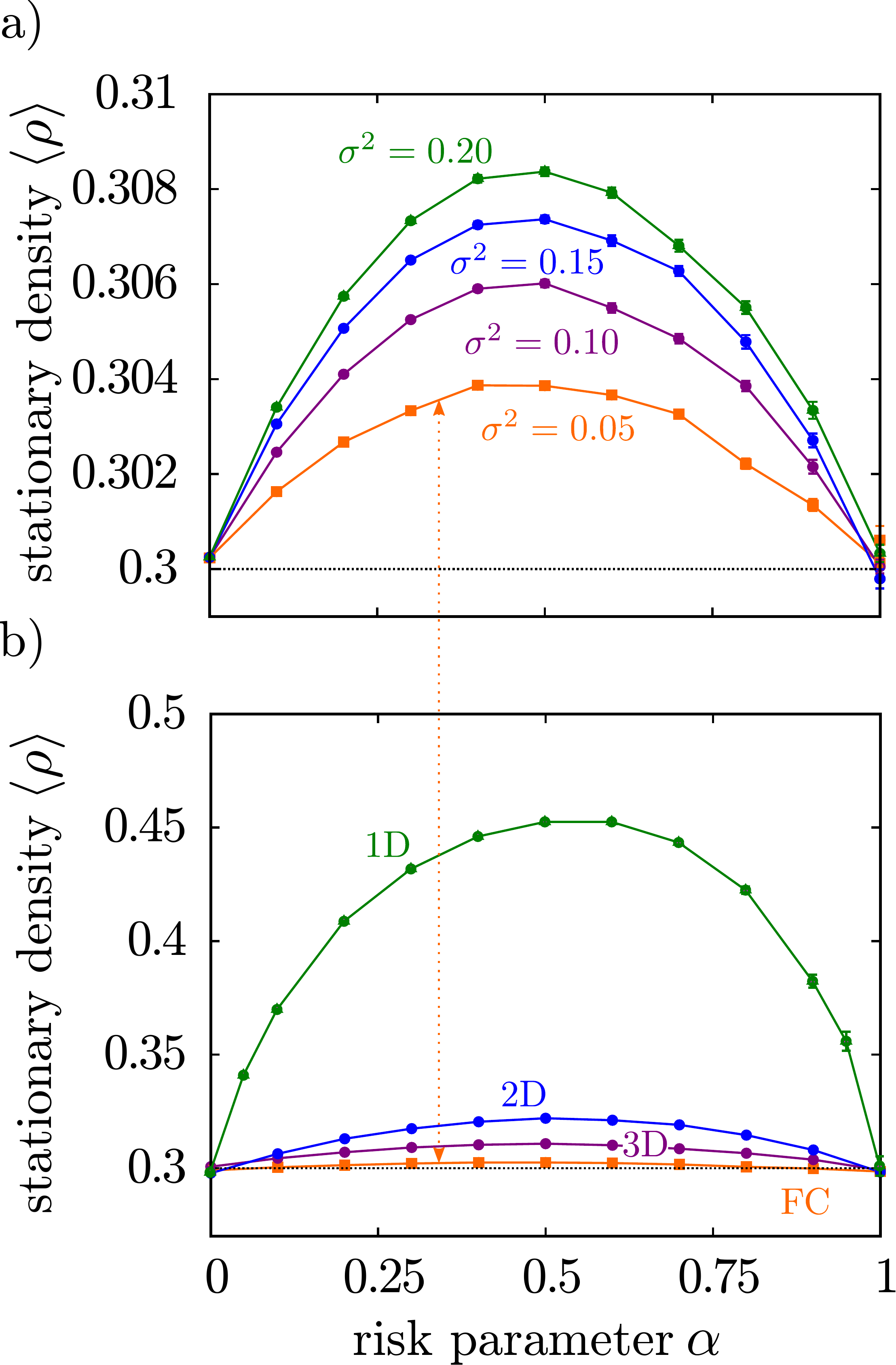} 
 \caption{ Effect (a) external-noise variability ($\sigma^2$) in the fully connected (FC) network
   and (b) dimensionality at fixed $\sigma^2$ on the stationary
   density for different bet-hedging strategies with asynchronous
   updating. Curves are results of Monte Carlo simulations of the CP with bet-hedging. 
   Qualitatively, the same
   conclusions are obtained as with the parallel updating. However,
   the benefits of bet-hedging strategies become lower in this second
   implementation, especially for the FC network.  Parameter values are as follows. (a) $N=10^4$;
   $p_0=0.588$; and $(\bar p,\sigma^2)$ of $(0.597,0.05)$,
   $(0.610,0.10)$, $(0.624,0.15)$ and $(0.637,0.20)$ in the different
   curves.  (b): $\sigma^2=0.05$; $N=10^4$ for 1D, 2D, and FC,
   $N=10648$ for 3D, and $(p_0,\bar p)$ are 1D:$(0.770,0.831)$,
   2D:$(0.655, 0.681)$, 3D:$(0.626,0.642)$ and
   FC:$(0.588,0.597)$. Most error-bars are smaller than symbol sizes. }
\label{fig:2noises-asynchronous}
\end{figure}

With this implementation, as in Fig. \ref{fig:2noises}, we have again
computed the curve $\ang{\rho(\alpha)}$ provided that
$\ang{\rho(0,1)}=0.3$, for different values of the external noise
variance $\sigma^2$ (in the FC network) and for different network
dimensions (fixing $\sigma^2$). As illustrated in
Fig. \ref{fig:2noises-asynchronous}, the relative position of all
curves is the same as for the case of synchronous updating: the
benefits of bet-hedging are enhanced as the noise amplitude is
increased. However, quantitatively, the enhancement is smaller than in
the synchronous case, discussed in the text.
 
As a final remark, observe that one could have naively expected
  that fluctuations derived from the sequential updating might contribute
  to an enhancement of the density for such hybrid strategies. This
  difference stems from the fact that in the sequential implementation
  of the model, not all individuals are necessarily updated at every
  single Monte Carlo step; thus the stochasticity introduced by this
  type of updating may save populations from extinctions in very
  unfavorable environments. This implies that the community does not rely as
  strongly on bet-hedging to perdure.

\section{Effect of temporal correlations}

A simple way to introduce temporal autocorrelations in the
environment is to take $p(t)$ to follow a Ornstein-Uhlenbeck process
\cite{gardiner}, i.e., a Brownian particle moving in a parabolic
potential. Mathematically, this process obeys \cite{gardiner}
\begin{equation}
\dot p = \theta (\bar p - p) + \sqrt{2\theta} \sigma \xi(t),
 \label{eq:ou}
\end{equation}
where $\bar p$ and $\sigma$ represent, as before, the mean and
variance of $p(t)$, respectively. With this choice, $p(t)$ is Gaussian
distributed, $N(\bar p,\sigma^2)$. The new parameter $\theta$ controls
the temporal autocorrelations, as $\ang{\left(p(t)-\bar
    p\right)\left(p(t')-\bar p\right)} = 
\sigma^2 e^{-\theta |t-t'|}$; consequently, $\theta\rightarrow0$ and
$\theta\rightarrow\infty$ represent the extreme cases of immutable
(completely correlated) and delta-correlated environments,
respectively. 

Equation (\ref{eq:ou}) can be integrated exactly, allowing for
a recursive generation of values at successive time steps \cite{toral-review},
\begin{equation}
p(t+1) = \bar p (1-e^{-\theta}) + p(t) e^{-\theta} + \sigma \sqrt{1-e^{-2\theta}} N(0,1),
 \label{eq:ou-integration}
\end{equation}
where $N(0,1)$ is a zero-mean unit-variance Gaussian random number.

\begin{figure}[b]
\centering  \includegraphics[width=\columnwidth]{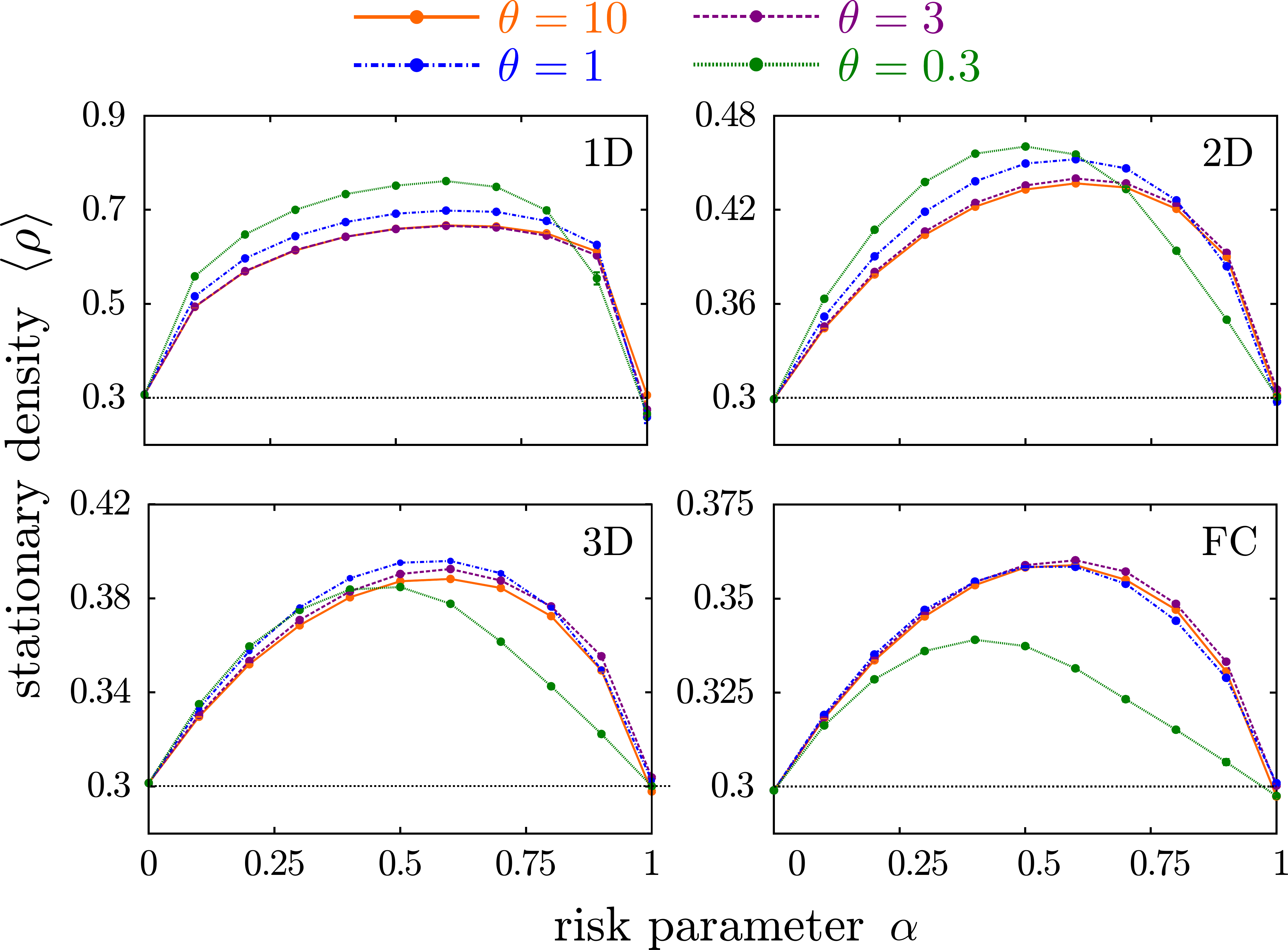}
\caption{Stationary density as a function of the risk parameter for
  different lattice topologies in temporal-correlated environments:
  the spreading probability of the risky strategy, $p(t)$ now obeys an
  Ornstein-Uhlenbeck process with mean $\bar p$, variance $\sigma^2$,
  and exponential temporal correlations with a characteristic time
  ${\theta}^{-1}$. The benefits of the hybrid strategies in the
  stationary density become enhanced for intermediate values of
  $\theta$. This result is much intensified in lower dimensions, while
  it is imperceptible for the 3D and FC networks. Parameters:
  $\sigma^2=0.05$; $(\bar p, \theta)$ in 1D: $(10,0.848)$,
  $(3,0.849)$, $(1,0.876)$, $(0.3, 0.944)$; 2D: $(10,0.704)$,
  $(3,0.706)$, $(1,0.719)$, $(0.3, 0.745)$; 3D: $(10,0.665)$,
  $(3,0.667)$, $(1,0.674)$, $(0.3, 0.681)$ and FC: $(10,0.628)$,
  $(3,0.629)$, $(1,0.631)$, $(0.3, 0.624)$; $p_0$ and $N$, taken as
  in Fig. \ref{fig:2noises}(b).}
\label{fig:ou}
\end{figure}

Fixing the environmental variance $\sigma^2$, we numerically
  study the effect of temporal correlations on bet-hedging for
different values of $\theta$ in every dimension. Following the same
strategy as above, we tune the parameters $p_0$ and $\bar p$ for each
temporal autocorrelation $\theta$ to fix the stationary density at
$\ang{\rho(\alpha=0,1)}$, and measure $\ang{\rho(\alpha)}$. Results
are summarized in Fig. \ref{fig:ou}.  Some remarks are in order. (i) The optimal strategy is always a hybrid strategy between
$\alpha=0$ and $\alpha=1$. Additionally, curves coincide with those in
Fig. \ref{fig:2noises}(b) when $\theta$ is high ($\theta\sim 10$), as
the external environment exhibits only short correlations. (ii)
When $\theta$ decreases moderately, the stationary density at the
optimal strategy becomes larger compared to the noncorrelated case.
In other words, bet-hedging strategies are more efficient for
temporally autocorrelated environments. This effect is stronger for
lower dimensions, whereas it barely applies to higher dimensional
lattices. (iii) When the autocorrelation is very large
($\theta<0.1$), the benefits of hybrid strategies are reduced compared
to the noncorrelated case, but they always remain convenient with
  respect to pure strategies. However, we are not interested in this
scenario, in which external conditions remain almost unchanged during
extremely long periods of time, and thus it behaves, effectively as a
constant-$p$ case. The inflection point in $\theta$ at which this
effect appears varies for different dimensions. (iv) Finally, the
optimal strategy becomes more conservative when temporal correlations
are added to the environment, with a bias to $\alpha^*\rightarrow 0$
when $\theta$ decreases. It would be nice to have a more detailed
analytical understanding of all this phenomenology, but we leave this challenging task for future work.

\begin{acknowledgements}  
We are grateful to R. Rubio de Casas for illuminating discussions and
to P. Moretti for a critical reading of the manuscript.  We
acknowledge support from J. de Andaluc{\'i}a Grant No. P09-FQM-4682 and
Spanish MINECO Grant Nos. FIS2012-37655-C02-01 and FIS2013-43201-P. 
\end{acknowledgements}

% \section*{References}
\bibliographystyle{apsrev}
%\bibliographystyle{unsrt}
% \bibliography{biblio-mixed-strategies}

\end{document}